\documentclass[aps, pra, twocolumn, showpacs,footinbib,superscriptaddress,longbibliography]{revtex4-1}
\usepackage{graphicx, subfigure,braket}
\usepackage{amsmath,amssymb}
\usepackage{color}
\usepackage{natbib,hyperref}
\usepackage{placeins}
\usepackage{appendix}
\usepackage[T1]{fontenc}
\usepackage{booktabs}
\usepackage{siunitx} 
\begin{document}

\title{Silicon-Germanium Heterostructures with Enhanced Valley Splitting for Spin Qubits}

\author{David W. Kanaar}
\affiliation{Department of Electrical and Computer Engineering,
University of California, Los Angeles, Los Angeles, CA 90095, USA}
\affiliation{Center for Quantum Science and Engineering, University of California, Los Angeles, Los Angeles, CA 90095, USA}
\author{Efrain Martinez}
\affiliation{Department of Physics, University at Buffalo, SUNY, Buffalo, NY 14260-1500, USA}
\author{Peihong Zhang}
\affiliation{Department of Physics, University at Buffalo, SUNY, Buffalo, NY 14260-1500, USA}
\author{Mark F.~Gyure}
\affiliation{Department of Electrical and Computer Engineering,
University of California, Los Angeles, Los Angeles, CA 90095, USA}
\affiliation{Center for Quantum Science and Engineering, University of California, Los Angeles, Los Angeles, CA 90095, USA}

\begin{abstract}
    Achieving valley splittings well in excess of the thermal energy of electrons and avoiding valley excitations is essential for the consistent initialization, operation and readout of gate-defined Si spin qubits. In this work, we present a device-level optimization strategy for pushing valley splittings to between 1 and 5 meV, well beyond values reported in nearly all previous theoretical studies. Using device-scale simulations that incorporate atomistic alloy disorder through a 1D tight-binding theory, we demonstrate that our proposed approach yields large valley splittings with a tight distribution across disorder realizations, a key requirement for reproducible qubit performance at scale. The approach rests on an unorthodox Si/SiGe heterostructure design combining a narrow quantum well, a small Ge spike, and a pure-Ge cap. We corroborate these predictions with targeted atomistic density functional theory calculations. These results offer a clear path forward for scalable Si/SiGe spin qubit devices and, if realized experimentally, effectively eliminate valley splitting as an existential problem for large scale SiGe-based quantum processors.
\end{abstract}
\maketitle

\section{Introduction}

Gate-defined Si/SiGe spin qubits are among the leading platforms for quantum computing because of their long coherence times, small on-chip footprint, and CMOS industry compatibility\cite{zwanenburg_silicon_2013,burkard_semiconductor_2023,neyens_probing_2024}. A persistent obstacle to scaling these devices is the small and alloy-disorder-sensitive nature of the valley splitting, the energy that  separates the two lowest conduction band states of a strained silicon quantum well. At typical operating temperatures, electron thermal energies are $k_B T \approx 15 \mu$eV. Qubit readout, gate operations, and coherent spin shuttling demand valley splitting significantly higher than this scale. Equally important, the random-alloy nature of the quantum well interfaces makes it so that even with large mean valley splitting, large variance can lead to unacceptably low yield in a large scale processor. Therefore we require not only a large average valley splitting but also a low relative variation across large areas.

A longstanding goal in this field has been to find heterostructures that have consistent valley splitting of order the orbital energy spacing (few meV), with low variations that would effectively eliminate valley splitting as an issue for qubit operation. Theoretical work has shown valley splitting in Si/SiGe depends on the Fourier weight of the potential, dominated by the alloy disorder, at 2$k_0$ that couples the two valleys. Several heterostructure designs have been proposed in this spirit: Ge spikes \cite{thayil_optimization_2025,losert_practical_2023,mcjunkin_sige_2022}, periodic Ge doping in the well\cite{mcjunkin_sige_2022,woods_coupling_2024,cvitkovich_increasing_2026}, sharp interfaces\cite{losert_practical_2023,paquelet_wuetz_atomic_2022}, and pure-Ge capping layers \cite{neyens_critical_2018,zhang_genetic_2013}. Except in highly idealized scenarios featuring non-smooth Ge profiles or significant Ge content within the well, no effective mass (EM), tight binding (TB) or density functional theory (DFT) results have shown consistent > 1 meV valley splitting across realizations. 

In this work we show theoretically that a Si/SiGe heterostructure built on an unorthodox combination of three design choices, a narrow quantum well between 2.5 and 3.0 nm, a thin pure-Ge cap in place of the conventional 30\% SiGe cap, and a dilute pure-Ge spike placed where the wavefunction envelope is largest, can deliver valley splittings as high as $\sim$5 meV with a narrow spread across alloy realizations, provided its key interfaces are atomically sharp. Because real interfaces are never perfectly abrupt, we then quantify how this ceiling erodes as the cap and spike interfaces are broadened, and find that the sharpness needed to retain a consistent valley splitting of $\sim$1 meV appears to be within the experimentally realizable limits. We argue that both the three design choices and these interface-width requirements are realistic targets for future optimized growth.

Our first design choice is adding a very thin pure-Ge cap before the 30\% SiGe buffer, which should present a much sharper top interface to the silicon well leading to higher valley splitting. A related structure, a $\sim$5 ML Ge layer atop a wider (13 nm) quantum well, has been studied experimentally via magnetotransport. The presence of the Ge layer in those measurements produced only a modest change in the measured valley splitting relative to a conventional interface but did not change mobility significantly\cite{neyens_critical_2018}. Pure-Ge interfaces of this kind were also studied in the context of Ge wetting layers and strain-driven (Stranski-Krastanov) island growth\cite{williams_strain_1991,liu_effect_1997}, where interdiffusion over only around 1-2 ML was observed, appreciably sharper than a typical 30\% SiGe-on-Si interface. This sharpness has a simple thermodynamic origin. The driving force for Si-Ge interdiffusion at the top of the well is set partly by the energetic preference for Ge-Ge nearest-neighbor bonding which is suppressed by a pure-Ge cap. This aligns with the abruptness of the experimental Ge wetting which was not optimized\cite{williams_strain_1991}. Additionally, the interface between the pure-Ge cap and the 30\% SiGe barrier above need not be sharp; this does not change the valley splitting since the electron wavefunction has decayed strongly by that depth. We will show that the use of a thin Ge cap by itself already dramatically increases the valley splitting over the values obtained with typical Si/SiGe interfaces.

However, the use of the Ge cap does not preclude combining this with other methods known to increase valley splitting. Our second design choice is a dilute Ge spike placed at the center of the well, where the wavefunction envelope is large.
The same energetic‑bonding argument suggests that we should see a pronounced spike at low Ge concentrations. Furthermore, we show that, somewhat surprisingly, even a small Ge fraction produces a substantial valley-splitting enhancement as long as the spike is sharp. This is an important finding, because introducing Ge into Si quantum wells has undesirable effects such as introducing disorder \cite{stehouwer_engineering_2025} to the potential landscape and increasing spin-orbit coupling \cite{woods_spin-orbit_2023}. Therefore it is desirable to find strategies that require as low Ge concentrations as possible. 

The final design choice is a thin quantum well. It is already well established \cite{chen_detuning_2021,losert_practical_2023} that valley splitting increases with decreasing quantum well width. In our proposed design using thin quantum wells, a large portion of the wavefunction interacts with both the pure-Ge cap and the Ge spike, additively enhancing the valley splitting. Silicon quantum wells as thin as 3 nm have already been grown and can host electrons with standard gate layouts and device designs\cite{chen_detuning_2021,acuna_coherent_2024}. Pushing to still thinner wells would yield consistently large valley splittings, but would face other design limits.

We found the optimized values for this heterostructure design by using atomically sharp interfaces and varying the pure-Ge cap thickness $N_\text{ML}$, well width $W_w$, and the Ge spike fraction $f_\text{Ge}$, identifying $N_\text{ML}$=1
$f_\text{Ge}$=0.04, and $W_w=$2.5-3.0 nm as optimal.
Using these optimized values we varied the sharpness of all features quantified by the spike and cap interface widths, $\sigma_\text{cap}$, and  $\sigma_\text{spike}$, and found at 1 ML or 2 ML the valley splitting $E_v \gtrsim 1\,\text{meV}$ is possible with a 3 nm and 2.5 nm well width respectively. 

\section{Results}\label{sec:results}

This section is organized in two parts. Section~\ref{sec:results:perfect} establishes an upper bound on the valley splitting attainable with our proposed quantum well design consisting of a narrow well, a thin Ge cap, and a Ge spike. By sweeping three independent parameters:  cap thickness, $N_{\mathrm{ML}}$, well width $W_w$, and spike Ge fraction $ f_{\mathrm{Ge}}$, using atomically sharp interfaces the optimal point for valley splitting is found. Section~\ref{sec:results:realistic} then quantifies how broadening the sharp cap and spike interfaces with a width of $\sigma_{\mathrm{cap}}$ and $\sigma_{\mathrm{spike}}$ decreases the valley splitting away from that ceiling, and identifies the sharpness profile required to retain a large valley splitting $E_v \gtrsim 1$~meV recently reported in Ref.~\cite{team_digitally_2026}.

For every parameter configuration, we generate 100 random alloy realizations, providing sufficient statistics to characterize the resulting valley splitting distribution. Throughout this section, these distributions are shown as box-and-whisker plots in which the central line marks the median, the box spans the interquartile range (IQR), and the whiskers extend to the most extreme realization within 1.5 IQR of the box edges.
 
\subsection{Idealized interface: the valley splitting ceiling}
\label{sec:results:perfect}

\begin{figure}[h!]
    \centering
    \includegraphics[width=0.9\linewidth]{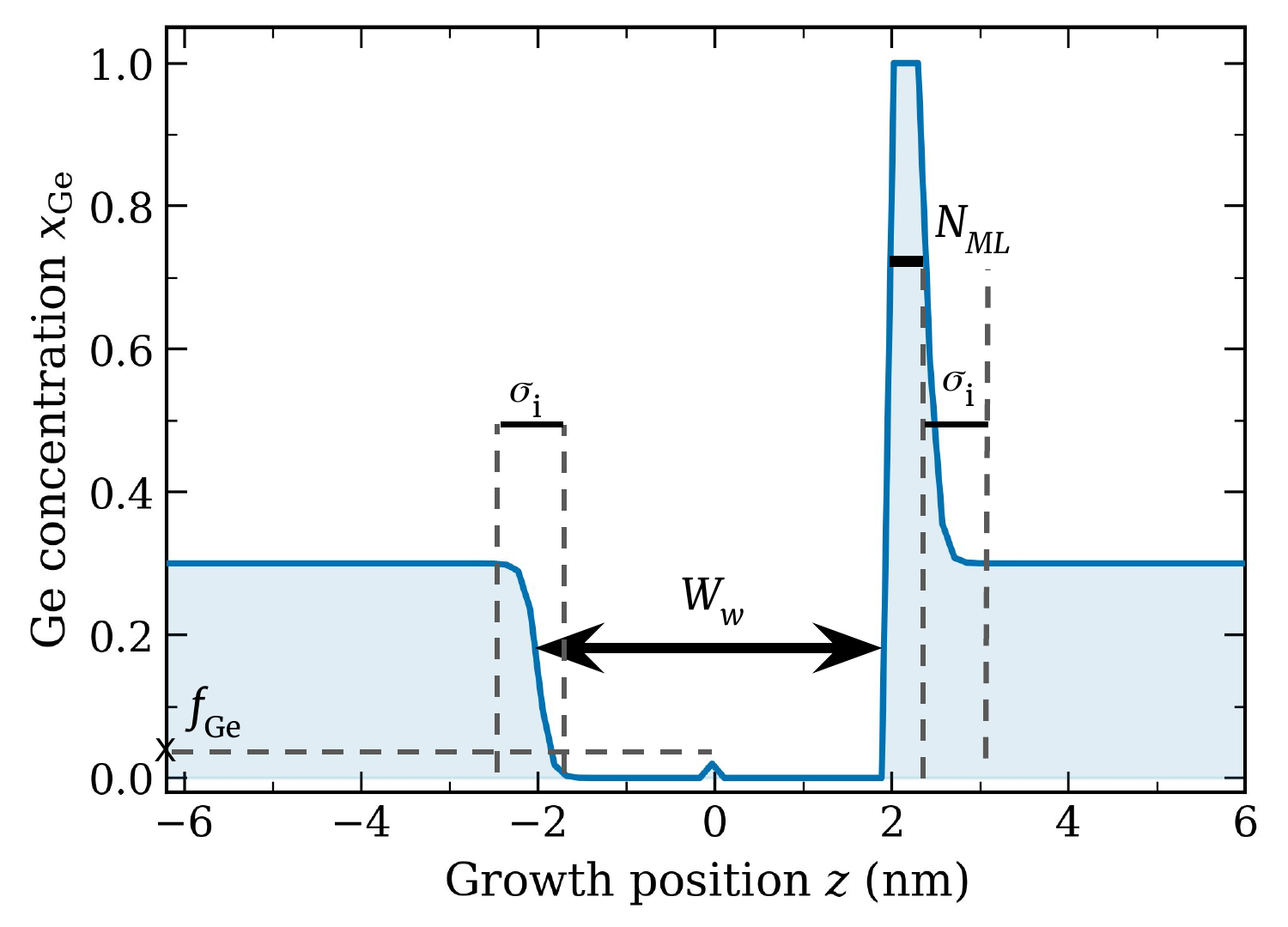}
    \caption{Ge profile used in the ideal interface case. The well width, $W_w$, number of cap monolayers $N_{\text{ML}}$, fraction of Ge used in the peak $f_{\mathrm{Ge}}$, and the sigmoid interface width for the non ideal interface $\sigma_i$ are all indicated.}
    \label{fig:GeProfileIdeal}
\end{figure}

We use a tight-binding inspired Hamiltonian equivalent to that in Ref.\cite{losert_practical_2023} and $2k_0$-theory to find the valley splitting as described in Appendix~\ref{sec:methods}. The idealized Ge concentration profile, which sets the probability of Ge in the random alloy disorder model, is shown in Fig.~\ref{fig:GeProfileIdeal}. Each of the three parameter sweeps reported below varies a single parameter around the baseline values in Table~\ref{tab:base_params}, chosen near the optimum identified in preliminary calculations. 

\begin{table}[h!]
  \centering
  \caption{Baseline parameters held fixed in the idealized interface sweep unless specified otherwise. Each sweep in Sec.~\ref{sec:results:perfect}
  varies a single parameter about these values. The orbital energy $\hbar \omega_i$ is swept in Sec.~\ref{app:result} while $F_z$ and $\sigma_i$ were not swept.}
  \label{tab:base_params}
  \begin{tabular}{l l c l}
    \toprule
    Parameter & Symbol & Value & Unit \\
    \midrule
    Pure-Ge cap thickness          & $N_{\mathrm{ML}}$        & $1$    & ML    \\
    In-plane $x$-confinement  \quad    & $\hbar\omega_x$          & $2.5$  & meV   \\
    In-plane $y$-confinement  \quad   & $\hbar\omega_y$          & $2.5$  & meV   \\
    Well width                     & $W_w$                    & $3.0$  & nm    \\
    Ge spike fraction              & $f_{\mathrm{Ge}}$        & $0.04$ &   \\
    Vertical electric field        & $F_z$                    & $5.0$ & MV/m  \\
    Interface sigmoid width \quad        & $\sigma_i$                    & $2.0$ & ML  \\
    \bottomrule
  \end{tabular}
\end{table}

\begin{figure}
    \centering
    \includegraphics[width=0.9\linewidth]{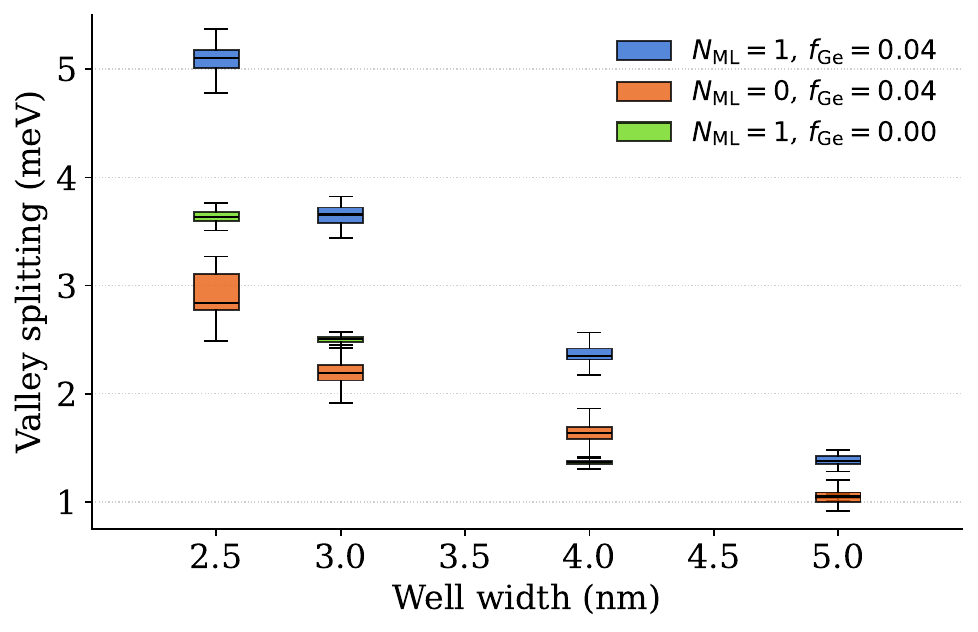}
    \caption{Distribution of the valley splitting over 100 alloy realizations as the well width $W_w$ is varied for $N_\mathrm{ML}=1$ and $f_\mathrm{Ge}=0.04$(blue),  $N_\mathrm{ML}=0$ and $f_\mathrm{Ge}=0.04$(orange), and  $N_\mathrm{ML}=1$ and $f_\mathrm{Ge}=0$(green) with the remaining parameters of Table \ref{tab:base_params}.}
    \label{fig:IdealWellwidth}
\end{figure}

The valley-splitting distribution as a function of well width, $W_w$, is shown in Fig.~\ref{fig:IdealWellwidth} with and without a cap. $N_{\text{ML}}=1$ is a one ML cap, zero means a sharp 30\% Ge interface. The Ge spike $f_{\text{Ge}}$ is $=0.04$ or $0$ as indicated. Valley splitting decreases substantially as well width increases as expected. Si quantum wells as thin as $W_w = 3$~nm have already been experimentally demonstrated\cite{chen_detuning_2021,acuna_coherent_2024}. Further well width reduction could push the valley splitting larger. However there are practical limits. One is set by the occupation of electron states at the SiO$_2$ interface; as the well becomes thinner the quantum well ground-state energy increases, and once it exceeds the energy of the states underneath the oxide they will fill before the well is occupied. Reducing the top SiGe barrier thickness brings the quantum well closer to the gates and partially restores its energetic preference. However, this also increases the coupling to oxide charge-noise defects\cite{connors_low-frequency_2019}. This trade-off between charge-noise sensitivity and the valley-splitting gain from thinning the well is out of the scope of this work and a 3.0 nm well will be considered a realistic number while 2.5 nm is considered within range for future optimization. Of note is that the mobility of electrons also reduces with decreasing well width. However, mobility is known to not directly impact qubit performance provided it is sufficiently large for electron transport through charge sensors and filling electron reservoirs. Further, including solely the spike or the cap still results in increased valley splitting as well width is decreased but combining both results in the largest effect.

\begin{figure}
    \centering
    \includegraphics[width=0.9\linewidth]{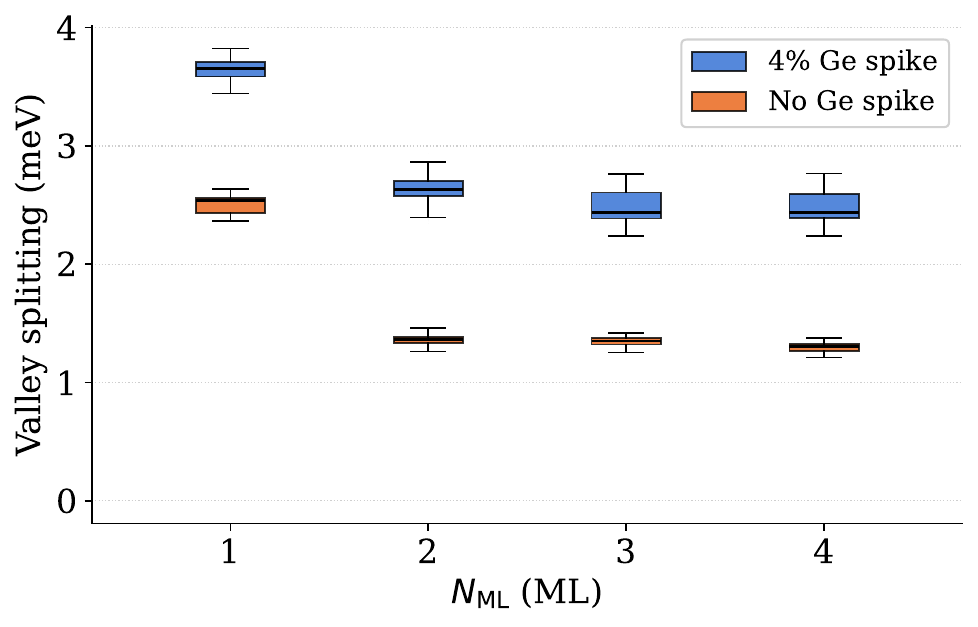}
    \caption{Distribution of the valley splitting over 100 alloy realizations as the cap thickness $N_{\mathrm{ML}}$ is varied about the base parameters of Table \ref{tab:base_params}(blue) and the base without the Ge spike(orange)}
    \label{fig:IdealNML}
\end{figure}

The valley-splitting distribution as a function of pure-Ge cap thickness, $N_{\mathrm{ML}}$, is shown in Fig.~\ref{fig:IdealNML} both with and without a 4\% Ge spike. The distributions are maximal at $N_{\mathrm{ML}}=1$ corresponding to Ge profiles in which the wavefunctions encounter a single sharp Ge feature. At the larger thicknesses, the valley wavefunctions decay in the cap but still sample the spread-out interfaces. The valley-coupling matrix element
from adjacent Ge layers partially cancel, suppressing the Fourier weight of the potential at the inter-valley wavevector $2k_0$~\cite{losert_practical_2023,ermoneit_exact_2026}. This is the same mechanism underlying the well-width dependence in Fig.~\ref{fig:IdealWellwidth}, sharper variations of the potential on a monolayer length scale couple more strongly the two valleys. Additionally, the dilute 4\% Ge spike is additive, increasing the valley splitting about 1 meV across different cap thicknesses. 
Pure Ge grown directly on Si is limited to the critical thickness of $\sim 3$-$4$~ML before a strain-driven Stranski-Krastanov transition produces 3D islands rather than layer-by-layer Ge growth~\cite{williams_strain_1991}. A pure-Ge cap deposited on a Si well that is pseudomorphically strained to the SiGe buffer, on the other hand, experiences a smaller lattice mismatch, and would plausibly have a larger critical thickness than $\sim 3$-$4$~ML when the underlying well is thin enough that little strain has relaxed. Among the realistic cap thicknesses considered, $N_{\mathrm{ML}} = 1$ yields the largest valley splitting, and we adopt this as the baseline throughout this work.

\begin{figure}
    \centering
    \includegraphics[width=0.9\linewidth]{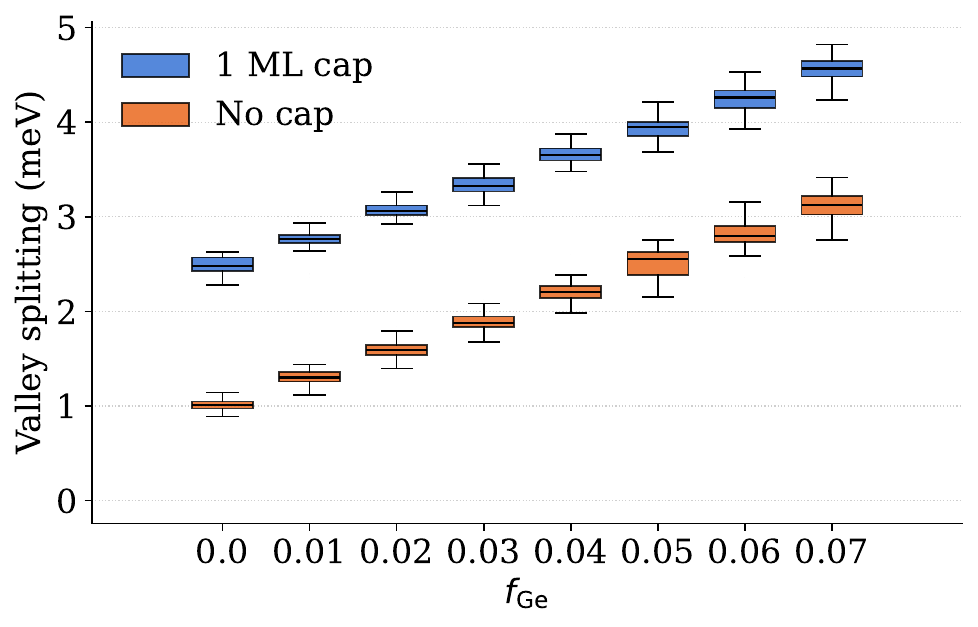}
    \caption{Distribution of the valley splitting over 100 alloy realizations as the Ge spike concentration $f_{\mathrm{Ge}}$ is varied about the base parameters of Table \ref{tab:base_params}.}
    \label{fig:IdealGeFrac}
\end{figure}

The valley-splitting distribution as a function of single-monolayer spike Ge fraction, $f_{\mathrm{Ge}}$, at $z=0$, is shown in Fig.~\ref{fig:IdealGeFrac} with and without a pure-Ge cap. The valley splitting grows approximately linearly with $f_{\mathrm{Ge}}$ over the tested range, so even a modest Ge content can substantially increase valley splitting. We adopt the small fraction of $f_{\mathrm{Ge}}=$ 4\% consistent with the design target for two reasons. First, Ge has stronger spin-orbit coupling than Si, and placing randomly distributed Ge atoms inside the well, where there is a large electron probability, also introduces spin-orbit coupling leading to relaxation channels detrimental to qubit operation~\cite{woods_spin-orbit_2023,losert_effects_2026,young_benchmarking_2025}. 
Second, a dilute peak is expected to be more stable against diffusion or segregation during growth. At low Ge content a spike atom sits in an otherwise pure-Si environment, so an exchange with a neighboring Si atom moves it between equivalent Si-only surroundings and gains no Ge-Ge bonding energy. Higher Ge fractions create more chances for the formation of Ge-Ge nearest-neighbor pairs which provide thermodynamic driving forces, promoting interdiffusion and smearing the spike. A dilute spike should therefore yield a sharper concentration profile than a more concentrated one. The same reasoning explained why a pure Ge cap should, in practice, be sharper than the conventional 30\% Ge cap.\par
We therefore adopt $f_{\mathrm{Ge}} = 4\%$ as the baseline for the remainder of the paper as this gives a >1 meV boost to the valley splitting assuming atomically sharp interfaces. At this percentage, the chance that one of the 8 atoms two bonds away is a Ge atom leading to diffusion is $1-(1-f_{\mathrm{Ge}})^8\approx27.9\%$ assuming all atoms in this region have the maximum 4\% Ge which is an overestimate.

\begin{figure}
    \centering
    \includegraphics[width=0.9\linewidth]{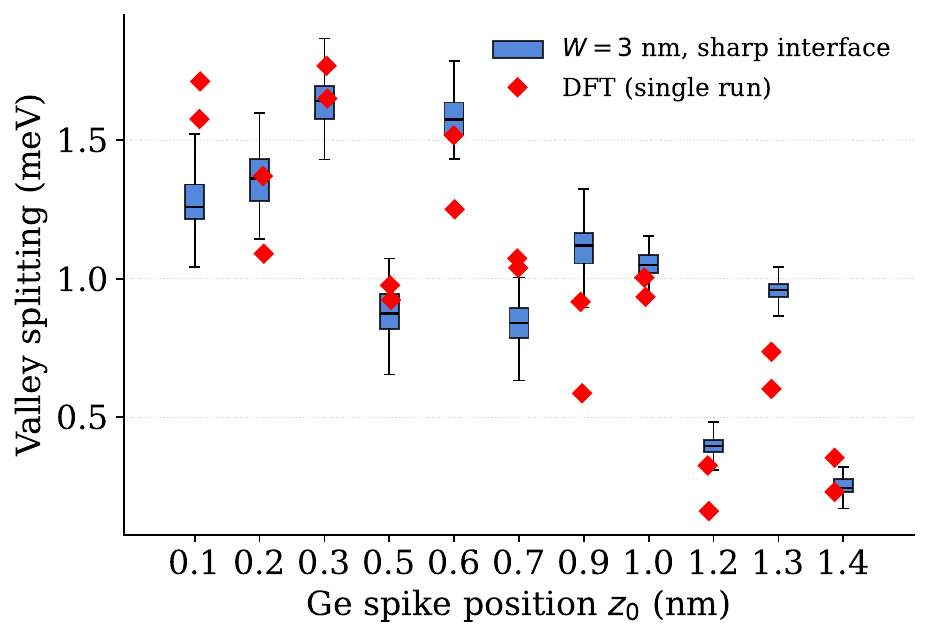}
    \caption{Distribution of the valley splitting over 100 alloy realizations using TB and two effective disorder realizations for DFT as the $f_{\mathrm{Ge}}=0.02$ Ge spike position is varied.}
    \label{fig:IdealGeSpikePos}
\end{figure}

To test whether the well center is the optimal spike position, we swept the Ge spike over all discrete monolayer sites and compared against DFT calculations of the same Ge profile. For this comparison we set $F_z=0$, $\sigma_i=0.01$ nm, and $N_{\mathrm{ML}}=0$ resulting in a sharp interface. The results are shown in Fig.~\ref{fig:IdealGeSpikePos}. The two DFT points per monolayer reflect the symmetry of the well about its center and follow the method in Appendix~\ref{sec:method:DFT}. Although the DFT and TB-modified EM methods are not directly comparable~\cite{cvitkovich_valley_2026}, they agree closely in magnitude and trend, both giving the largest valley splitting for a spike near the center of the well, where the wavefunction amplitude is largest. We therefore place the spike at $z=0$ as defined in Fig.~\ref{fig:GeProfileIdeal}.

Taken together, these sweeps show that the proposed geometry can in principle deliver valley splittings of up to $\sim 5$~meV, provided the well is narrow ($W_w \approx 2.5$~nm), the pure-Ge cap is a single monolayer, a small Ge spike is placed inside the well, and these interfaces are atomically sharp, even while other interfaces are diffused. The remaining question is how sharp the interfaces actually need to be for these splittings to survive in a realistic structure. This is addressed in the next section.

\subsection{Realistic interfaces}
\label{sec:results:realistic}
\begin{figure}
    \centering
    \includegraphics[width=0.9\linewidth]{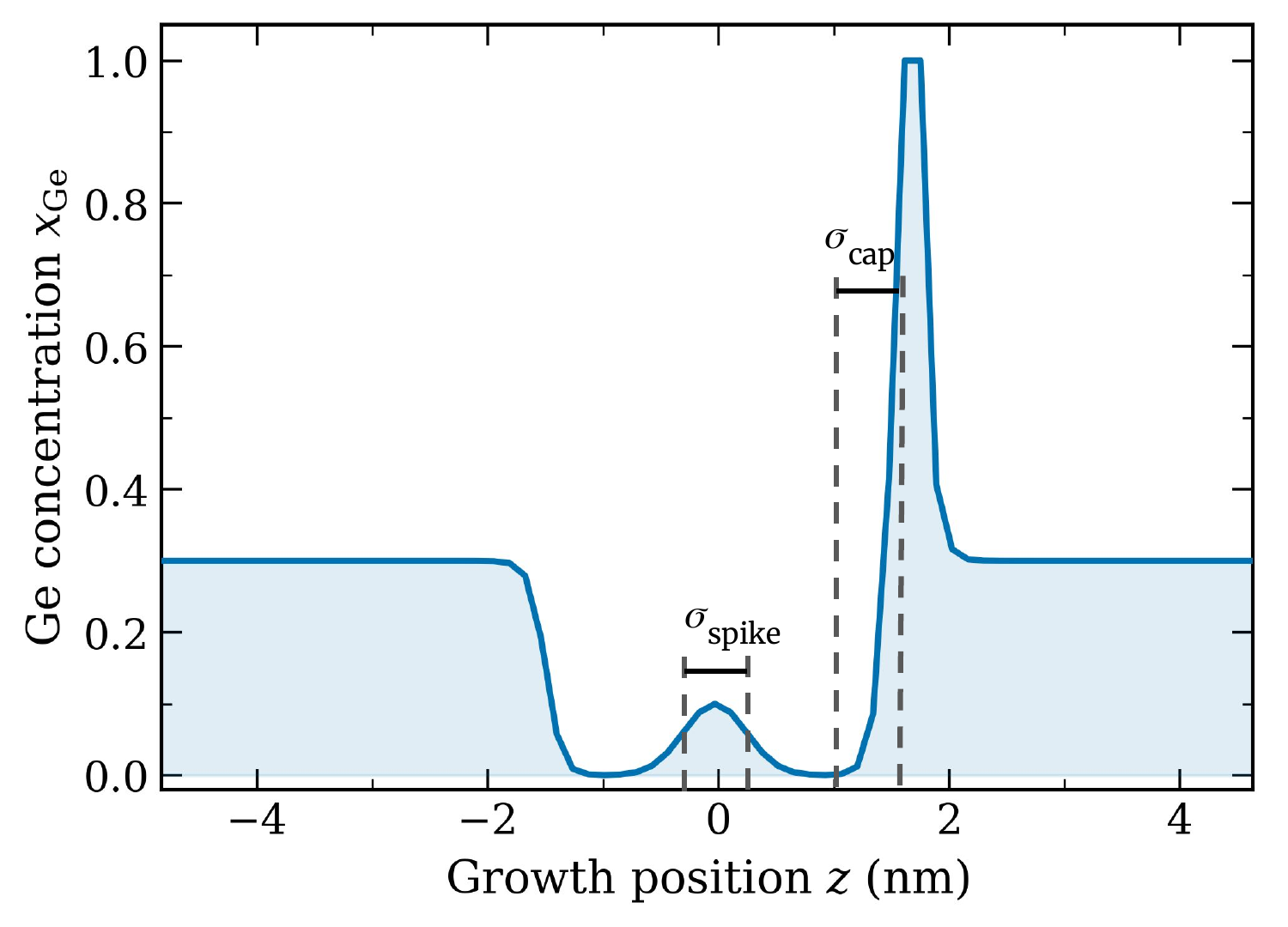}
    \caption{Ge profile used in the realistic interface case. The additional features, the cap sigmoid with interface for $\sigma_\mathrm{cap}$ and the spike Gaussian width $\sigma_\mathrm{spike}$ are indicated.}
    \label{fig:Geprofile2}
\end{figure}

Realistic heterostructure interfaces are never atomically sharp because Si-Ge intermixing during the growth broadens the Ge profile over a finite length scale. Therefore, we relax the sharp-interface assumption of Sec.~\ref{sec:results:perfect}, retaining the optimum geometries identified there, $W_w = 2.5$~nm or 3.0nm, $N_{\mathrm{ML}} = 1$, $f_{\mathrm{Ge}} = 4\%$. We do this by replacing the idealized Ge profile of Fig.~\ref{fig:GeProfileIdeal} with a finite-width transition shown in Fig.~\ref{fig:Geprofile2}. The Ge spike is modeled as a Gaussian of height $f_{\mathrm{Ge}}$ and width $\sigma_{\mathrm{spike}}$, and the Si-Ge boundary between the well and the pure-Ge cap as a sigmoid of width $\sigma_{\mathrm{cap}}$ with the full functional form given in Sec.~\ref{sec:methods}.

\begin{figure}
    \centering
    \includegraphics[width=0.9\linewidth]{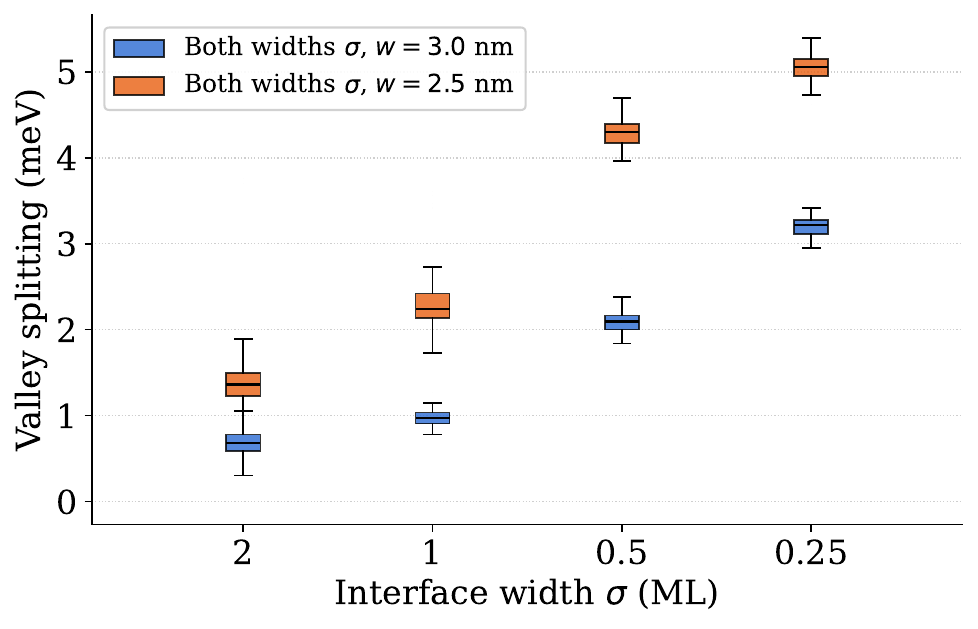}
    \caption{Distribution of the valley splitting over 100 alloy realizations as the cap width $\sigma_\mathrm{cap}$ and the spike width $\sigma_\mathrm{spike}$ are jointly varied for $W_w=3.0$nm well(blue), and $W_w=2.5$nm well(orange). The unspecified parameters are from Table \ref{tab:base_params}.}
    \label{fig:realisticVS}
\end{figure}

To identify the sharpness needed to retain a valley splitting of $E_v \gtrsim 1$~meV in this realistic setting, we report two sweeps in Fig.~\ref{fig:realisticVS}. We sweep the interface widths $\sigma_{\mathrm{cap}}$ and \ $\sigma_{\mathrm{spike}}$ jointly at a well width of $W_w=$3.0 nm and a well width of $W_w=$2.5 nm. \par
A valley splitting of $E_v \gtrsim 1$~meV is achieved when both $\sigma_{\mathrm{cap}}$ and $\sigma_{\mathrm{spike}}$ are $\sim 1$~ML and the well width $W_w=3.0$nm or both interface widths are $\sim 2$~ML and the well width $W_w=2.5$ nm. The disorder spread at $\sigma_{\mathrm{cap}}=\sigma_{\mathrm{spike}}=1$~ML and $W_w=3.0$ nm is relatively narrow with a standard deviation of $\sim 190\mu$~eV  and a $E_v \gtrsim 1$~meV while the 2 ML interface width and 2.5 nm well width have no data points below 1 meV. The practical conclusion is that producing $E_v \gtrsim 1$~meV requires only one heterostructure improvement above current published state-of-the-art devices, a 2.5 nm well width or 1 ML sharp cap and spike interfaces.
\begin{figure}
    \centering
    \includegraphics[width=0.8\linewidth]{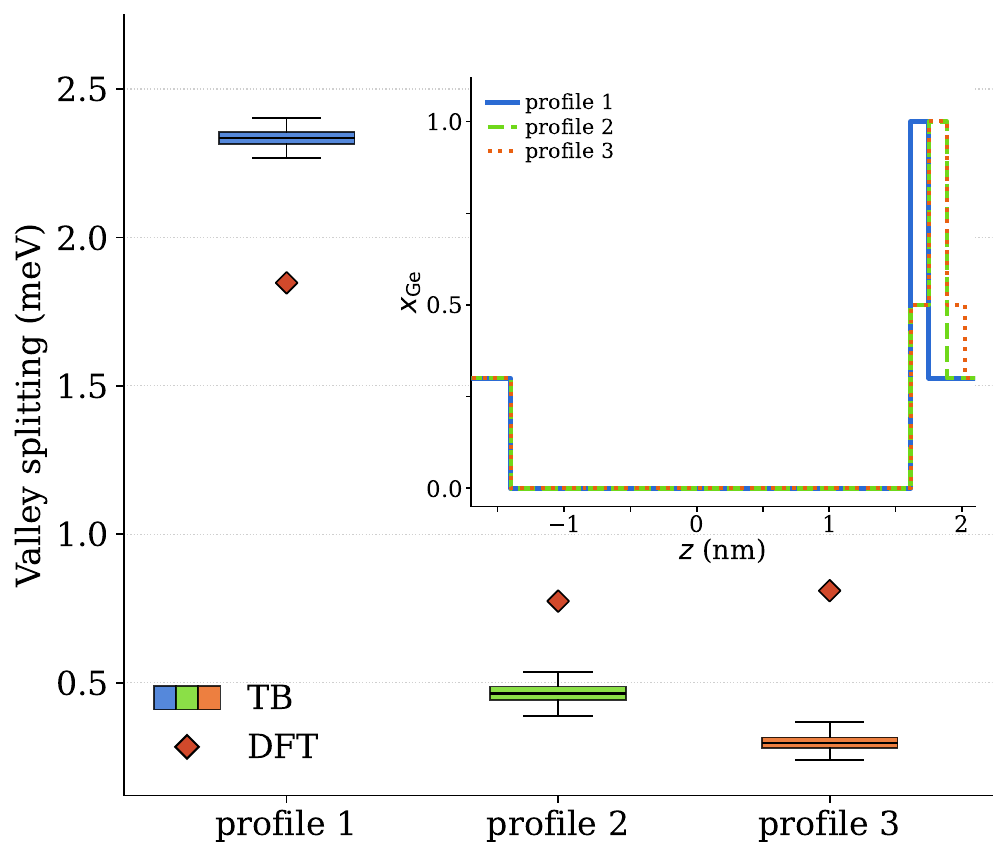}
    \caption{Valley splitting for three different Ge profiles, computed with two methods. For the TB method, box-and-whisker plots show the distribution over 100 disorder realizations. For DFT, points mark the valley splitting of each individual structure tested. Inset: the Ge profiles used to compare the DFT and simplified TB methods.}
    \label{fig:DFTcompare}
\end{figure}
To check our methods and confirm the trends we observe survive in a more atomistic treatment, we compare to density functional theory (DFT) calculations, described in Sec.~\ref{sec:method:DFT}. The two methods do not solve the exact same problem and making them quantitatively agree requires strict tuning~\cite{cvitkovich_valley_2026}. The trend they share, however, is expected to be the same and would confirm whether a sharp Ge cap increases the valley splitting. To this end, we compare three Ge profiles shown in the inset of Fig.~\ref{fig:DFTcompare}. The first has a sharp pure-Ge cap of 1 ML. The second adds a monolayer with 50\% Ge below the cap. The third adds a monolayer with 50\% Ge below and above the cap. Both methods use the same Ge profile, however, for the TB method the valley splitting is extracted for 100 disorders per profile, while for DFT we construct a smaller cell of atoms with these profiles and extract the valley splitting as described in Sec.~\ref{sec:method:DFT}. The results for both methods are shown in Fig.~\ref{fig:DFTcompare}. The two methods reveal the same trend. The low valley splitting in the TB results highlights the need for even a low-percent Ge spike to push the splitting to 1 meV. This shared trend confirms that interface sharpness is the controlling factor, and that our methods capture it correctly.

\section{conclusion}
We have proposed an unorthodox Si/SiGe heterostructure that combines three design choices: a narrow quantum well, a single-monolayer pure-Ge cap in place of the conventional 30\% SiGe cap, and a dilute Ge spike at the center of the well. With atomically sharp interfaces, this design yields valley splittings in excess of 5 meV, far above the values reported in nearly all previous theoretical work, and, just as importantly, with a tight distribution across alloy-disorder realizations. Because real interfaces are never perfectly abrupt, we quantified how this ceiling erodes as the cap and spike are broadened. We find that a consistent valley splitting of $E_v \gtrsim 1$ meV survives for interface widths of ~1 ML at a 3.0 nm well and ~2 ML at a 2.5 nm well, sharpnesses that appear to lie within experimental reach, and that we argue are thermodynamically favored for a dilute spike and a pure-Ge cap. Our DFT calculations reproduce the same trend, confirming that interface sharpness is the controlling factor. Taken together, these results show that valley splitting need not remain an obstacle to scaling Si spin qubits: a modest improvement in well width and interface sharpness over current devices would effectively eliminate it as a barrier to large-scale Si/SiGe spin-qubit processors.

\section{acknowledgments}
  The primary in-house software used for this work was Chris Anderson's MaSQE components and the MaSQE FCI programs; we would like to thank him for both the use of, and assistance in, modifying this software necessary to create the results presented in this paper. This work used computational and storage services associated with the Hoffman2 Cluster which is operated by the UCLA Office of Advanced Research Computing’s Research Technology Group. PZ acknowledges computational
support from the Center for Computational Research, University at Buffalo, SUNY. The authors acknowledge support from the Air Force Office for Scientific Research (AFOSR) under Grant Number FA9550-23-1-0710. 

\bibliographystyle{apsrev4-1}
\bibliography{refs}

\newpage
\newpage
\appendix
\newpage

\section{Methods}
\label{sec:methods}

We compute the valley splitting within a single-band effective-mass description that reproduces the two silicon $\Delta_z$ conduction-band minima at $k_z=\pm k_0$. The calculations are equivalent to a two-band tight-binding model in the $z$-direction\cite{boykin_valley_2004,losert_practical_2023} and are referred to as $2k_0$ theory\cite{cvitkovich_valley_2026,ermoneit_exact_2026}. Compared to full 3D tight-binding calculations, this approach trades atomistic faithfulness for speed. The Hamiltonian is evaluated on a rectangular real-space grid while the kinetic operator is applied in $k$-space using the fast Fourier transform (FFT). The lowest four eigenstates are obtained through a Rayleigh-Chebyshev iteration\cite{anderson_rayleighchebyshev_2010} and the valley splitting is extracted as the energy difference between the lowest non-orbital excited states. The method as employed in this manuscript solely aims to quickly estimate the eigenstate energies and the details of the wavefunction are not important. Strain effects beyond the valley minima being at $k_z=\pm k_0=0.82\times 2 \pi/a_0$ are ignored. Although strain engineering is possible it is more difficult in practice and we focus on alloy-disorder engineering with existing stack structures\cite{woods_coupling_2024}.   

\subsection{Single-particle Hamiltonian}
The single-particle Hamiltonian takes the form
\begin{equation}
    H_{\text{sp}}(\mathbf{r}) = H_{\text{kin}}(\mathbf{r}) + V(\mathbf{r}),
    \label{eq:Hsp}
\end{equation}
where $H_{\text{kin}}$ is the effective-mass kinetic operator and $V(\mathbf{r})$ is the position-dependent potential.

The kinetic operator is calculated in momentum space, where the two valleys appear as parabolic minima at $k_z = \pm k_0$ with $k_0 = 0.82 \times 2\pi/a_0$ and $a_0$ the silicon lattice constant,
\begin{equation}
H_{\text{kin}}(\mathbf{k})
 = \frac{\hbar^2 (k_x^2 + k_y^2)}{2 m_{xy}}
 + \begin{cases}
       \dfrac{\hbar^2 (k_z - k_0)^2}{2 m_z} & k_z \ge 0 \\
       \dfrac{\hbar^2 (k_z + k_0)^2}{2 m_z} & k_z < 0,
   \end{cases}
\label{eq:Hkin}
\end{equation}
where $m_{xy}$ and $m_z$ are the transverse and longitudinal conduction-band effective masses.  $H_{\text{kin}}$ is applied with a split-operator method efficiently using FFTs.

To make sure the spectral kinetic operator is converged we refine the grid along $z$ finer than the underlying atomic monolayer spacing, so that the $\pm k_0$ minima are well within the first Brillouin zone of the grid.  We therefore sample the wavefunction on a $z$-grid with spacing $\Delta z=a_0/12$. This fine grid is reconciled with the atomic alloy distribution grid discussed in Sec.~\ref{sec:methods:VCA}.

The confining potential has three pieces: a parabolic in-plane confinement, a uniform vertical electric field, and a position-dependent conduction-band offset from the SiGe alloy,
\begin{equation}
    V(\mathbf{r})
      = \tfrac{1}{2} m_{xy}\, \omega_x^2\, x^2
      + \tfrac{1}{2} m_{xy}\, \omega_y^2\, y^2
      + e F_z\, z
      + V_{\text{bo}}(\mathbf{r}),
    \label{eq:V}
\end{equation}
where $\hbar\omega_{x,y}$ set the lateral orbital energies, $F_z$ is the vertical electric field, and $V_{\text{bo}}(\mathbf{r})$ is the alloy-disorder conduction-band offset described next.

\subsubsection{Random alloy and virtual-crystal band offset}
\label{sec:methods:VCA}

The alloy disorder of the underlying diamond-lattice grid is generated on a rectangular grid with in-plane spacing $d_x = d_y = a_0/\sqrt{2}$ and vertical spacing $d_z = a_0/4$ between silicon monolayers~\cite{losert_practical_2023}. Each lattice site is assigned an occupation $n_{\text{Ge}}(\mathbf{r}_a) \in \{0,1\}$ drawn from an independent Bernoulli distribution with position-dependent probability $\chi_{\text{Ge}}(z_a)$. The vertical profile $\chi_{\text{Ge}}(z)$ encodes the heterostructure design: a SiGe barrier of fixed Ge fraction $\chi_\mathrm{S} = 0.30$, a Si quantum well of width $W_w$ centered at $z = 0$, an optional Ge spike of fraction $f_\mathrm{Ge}$ at $z = 0$, and a pure-Ge cap of $N_\mathrm{ML}$ monolayers above the well, with sigmoidal interfaces of width $\sigma_i$ at each region boundary.

Defining $z_c \equiv W_w/2 + N_\mathrm{ML}\,a_0/4$ as the top of the cap, the profile is
\begin{equation}
\chi(z) =
\begin{cases}
    f_\mathrm{bot}(z) & z < \dfrac{W_w}{2}, \\[8pt]
    1                 & \dfrac{W_w}{2} \le z < z_c, \quad N_\mathrm{ML} > 0, \\[8pt]
    f_\mathrm{top}(z) & z \ge z_c, \quad N_\mathrm{ML} > 0, \\[8pt]
    \chi_\mathrm{S}   & z \ge \dfrac{W_w}{2}, \quad N_\mathrm{ML} = 0,
\end{cases}
\label{eq:chiz}
\end{equation}
with sigmoid functions
\begin{align}
f_\mathrm{bot}(z) &= \frac{\chi_\mathrm{S}}{1 + \exp\!\left(\dfrac{4\,(z + W_w/2)}{\sigma_i}\right)},
\label{eq:fbot} \\[6pt]
f_\mathrm{top}(z) &= \frac{1 - \chi_\mathrm{S}}{1 + \exp\!\left(\dfrac{4\,(z - z_c)}{\sigma_i}\right)} + \chi_\mathrm{S}.
\label{eq:ftop}
\end{align}
The optional Ge spike is included by overriding the profile at the monolayer nearest to $z = 0$,
\begin{equation}
\chi(0) = f_\mathrm{Ge}.
\label{eq:spike}
\end{equation}

Two modifications generalize the profile of Eq.~\eqref{eq:chiz} to a non-sharp interface. First, the Ge spike at depth $z_0$ is broadened from a single monolayer into a Gaussian,
\begin{equation}
    g(z) = f_\mathrm{Ge}\,\exp\!\left(-\frac{(z-z_0)^2}{2\sigma_{\mathrm{spike}}^2}\right),
    \label{eq:spike_gauss}
\end{equation}
with amplitude $f_\mathrm{Ge}$ and width $\sigma_{\mathrm{spike}}$.
Second, the abrupt Si-to-cap transition is softened in the upper half of the well
($0 \le z < W_w/2$), a new sigmoid ramp,
\begin{equation}
    f_\mathrm{cap}(z) = \frac{1}{ 1+ \exp\ \left( \dfrac{4\,(W_w/2-z)}{ \sigma_\mathrm{cap} }   \right)},
    \label{eq:fcap}
\end{equation}
builds the Ge concentration smoothly from $\approx 0$ at $z=0$ to 1 at the pure-Ge cap ($z = W_w/2$), controlled by the cap-interface width $\sigma_{\mathrm{cap}}$.

The total non-sharp profile is
\begin{equation}
    \chi(z) =
    \begin{cases}
        f_\mathrm{bot}(z)   & z < 0, \\[8pt]
        f_\mathrm{cap}(z)   & 0 \le z < \dfrac{W_w}{2}, \\[8pt]
        1                   & \dfrac{W_w}{2} \le z < z_c, \quad N_\mathrm{ML} > 0, \\[8pt]
        f_\mathrm{top}(z)   & z \ge z_c,
    \end{cases}
    \label{eq:chiz_soft}
\end{equation}
with $f_\mathrm{bot}$ and $f_\mathrm{top}$ as in Eqs.~\eqref{eq:fbot}-\eqref{eq:ftop},
and the Gaussian spike added as $\chi(z) \to \chi(z) + g(z)$.
The Ge profiles used for the idealized and diffused cases are plotted in Sec.~\ref{sec:results}.

\subsubsection{Atomic-to-simulation grid mapping}

Because the simulation $z$-grid is finer than the atomic spacing, the wavefunction value at a simulation point generally lies between two adjacent monolayers.  Each simulation cell at
position $\mathbf{r}_i$ is assigned a Ge content
\begin{equation}
    X_{\text{Ge}}(\mathbf{r}_i)
       = (1 - s_z)\, \bar n_{\text{Ge}}(z_a^{(0)},  A_i)
         +  s_z   \, \bar n_{\text{Ge}}(z_a^{(1)},  A_i),
    \label{eq:interp}
\end{equation}
where $z_a^{(0)}$ and $z_a^{(1)}$ are the two atomic monolayers that surround $z_i$, $s_z \in [0,1]$ is the relative position of $z_i$ between
them, and $\bar n_{\text{Ge}}(z_a,  A_i)$ is the mean Ge occupation of the atomic sites of monolayer $z_a$ that fall within the in-plane footprint $A_i$ of simulation cell $i$. $A_i$ is $2\times2\,\,\, \text{nm}^2$ such that it contains multiple atoms. Note that $X_\text{Ge}$ refers to the Germanium concentration in a grid box which is distinct from $\chi(z)$ which refers to the concentration profile as a function of $z$. Eq.~\eqref{eq:interp} amounts to a linear interpolation of the virtual crystal approximation (VCA) Ge fraction in $z$ from the atomic monolayer grid onto the simulation grid.

Within the VCA, each simulation cell is assigned an effective Ge fraction equal to the average occupation of the atomic sites that fall within it. The conduction-band offset relative to the unstrained SiGe background follows the strained-on-SiGe expression of Ref.~\cite{schaffler_high-mobility_1997},
\begin{equation}
    \begin{aligned}
        \Delta E_c 
        &= (X_w - X_\mathrm{S}) \left[ \frac{X_w}{1 - X_\mathrm{S}} \bigl(-0.502(1 - X_\mathrm{S})\bigr) \right. \\
        &\quad \left. - \frac{1 - X_w}{X_\mathrm{S}} \bigl(0.743 - 0.625(1 - X_\mathrm{S})\bigr) \right],
    \end{aligned}
    \label{eq:schaffler}
\end{equation}
where $X_w$ and $X_\mathrm{S}$ are the Si fractions of the well and of the SiGe substrate that sets the strain reference\cite{schaffler_high-mobility_1997,losert_practical_2023}.  In the calculation we use $X_w= 1$ (pure Si well) and $X_\mathrm{S}= 0.70$ as references, the band offset that enters the Hamiltonian in each simulation cell is then linearly rescaled by the local Si content of that cell $X_{\text{Si}}(\mathbf{r}) = 1 - X_{\text{Ge}}(\mathbf{r})$, as
\begin{equation}
    V_{\text{bo}}(\mathbf{r})
       = -\,\Delta E_c\;
         \frac{X_{\text{Si}}(\mathbf{r}) - X_w}{X_\mathrm{S} - X_w},
    \label{eq:Vbo}
\end{equation}
where $X_{\text{Ge}}(\mathbf{r})$ is calculated using Eq.~\eqref{eq:interp}.
With this convention $V_{\text{bo}} = 0$ in a pure-Si cell, $V_{\text{bo}}$ equals the full SiGe-Si offset in a 30\% Ge cell, and $V_{\text{bo}}$ correctly extrapolates to the pure-Ge cells in the cap.

\subsection{DFT modeling}\label{sec:method:DFT} 

DFT calculations are carried out using the Vienna Ab-initio Simulation Package (VASP) \cite{kresse_ab_1993,kresse_efficiency_1996}. We use the PBEsol \cite{perdew_restoring_2008} exchange-correlation energy functional for the structural optimizations since this functional is known to be able to reproduce accurately
the structural parameters for solids. Periodic SiGe/Si/SiGe superlattice structures, with or without
additional Ge layers, are used to model Si QWs sandwiched between Si$_{0.7}$Ge$_{0.3}$ barriers. 
The Si$_{0.7}$Ge$_{0.3}$ random alloys are generated using the special quasi-random structure (SQS) 
method \cite{zunger_special_1990,walle_alloy_2002}. All structures are first optimized until the force on each atom is less than 0.002 eV/\AA. The optimized lattice constant for the Si$_{0.7}$Ge$_{0.3}$ random alloy is 
5.507 \AA~(rescaled for an 8-atom cubic cell), which compares well with the 
experimental value of 5.497 \AA~for a fully relaxed alloy ~\cite{dismukes_lattice_1964}. 
In the subsequent calculations, the lateral lattice constants of the Si/SiGe structure 
are fixed at 5.507 \AA, while the perpendicular lattice constant (particularly in the Si QW) is allowed to relax.

The two nearly degenerate $\pm z$-valleys are the lowest energy conduction band states 
with wave functions confined within the QW. 
We would like to mention that accurate calculations of the valley splitting of SiGe/Si/SiGe heterostructures using DFT can be challenging.
First, large Si$_{0.7}$Ge$_{0.3}$ random alloys are needed to better model
experimental structures. In this work, we generate several large random structures containing thousands of atoms (i.e., $5\times 5\times 12$ cubic cells).
Second, accurately resolving small valley splittings ($\sim$ 10 $\mu$eV) 
requires a very high level of convergence in the calculations. The eigenvalues must 
converge to $\lesssim$ 1 $\mu$eV and the structures must also be fully relaxed. 
During the process of structural relaxation, we sometimes observe over 1 meV fluctuation in the
calculated valley splitting. 

\section{Appendix: Results}\label{app:result}
\begin{figure}
    \centering
    \includegraphics[width=0.9\linewidth]{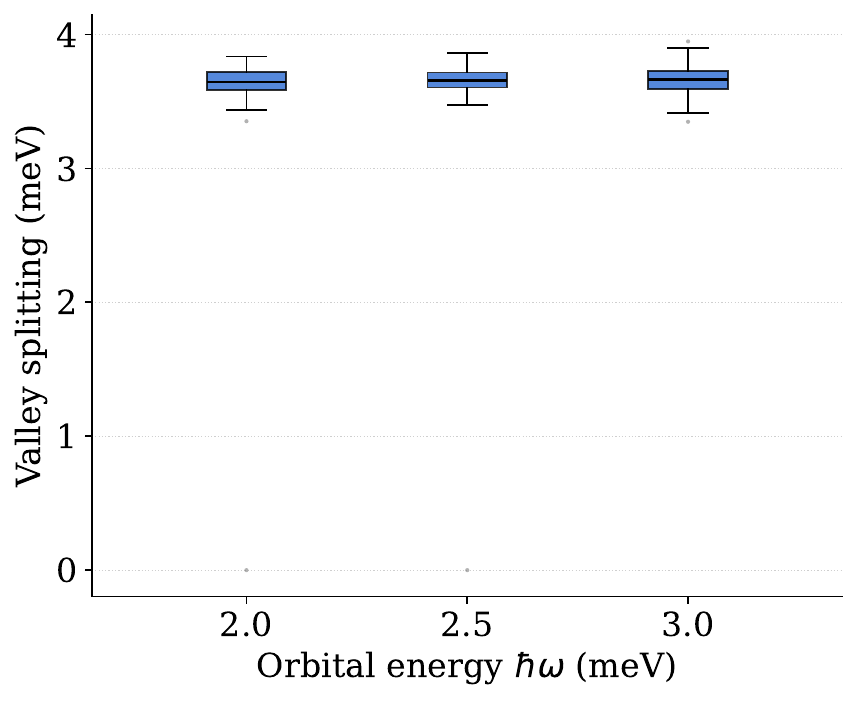}
    \caption{Distribution of the valley splitting over 100 alloy realizations as the orbital energy $\hbar \omega$ is varied about the base parameters of Table \ref{tab:base_params}. For each value, the central line marks the median, the box spans the interquartile range (IQR), and the whiskers extend to the most extreme realization within 1.5 IQR of the box edges.}
    \label{fig:IdealOrbital}
\end{figure}

The valley splitting distribution for different in-plane orbital confinement energies $\hbar\omega_x = \hbar\omega_y \equiv \hbar\omega$ is shown in Fig.~\ref{fig:IdealOrbital}. The valley splitting average does not change significantly across the swept range, however the variance does slightly increase with increased confining energy. This is mainly because the in-plane radius of the wavefunction is proportional to the inverse of the square root of the orbital energy $r\propto\sqrt{1/\omega}$. This means the radius does not change much for energies measured in devices. As a result the in-plane orbital energy, primarily set by the gate geometry and heterostructure, does not need to be tuned to maximize valley splitting. We do note increasing the orbital energy and thereby increasing dot radius could increase valley splitting if pushed sufficiently as in other works\cite{rahlff_sharp_2026}. \par
\begin{figure}
    \centering
    \includegraphics[width=0.9\linewidth]{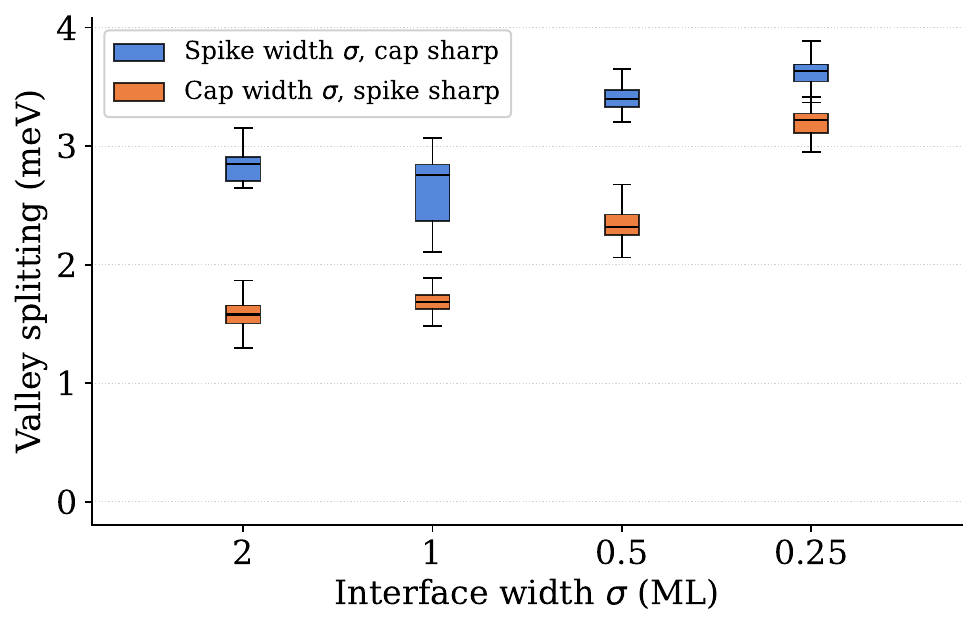}
    \caption{Distribution of the valley splitting over 100 alloy realizations as either the cap width $\sigma_\mathrm{cap}$ or the spike width $\sigma_\mathrm{spike}$ are varied while the other interface is kept atomically sharp. The unspecified parameters are from Table \ref{tab:base_params}.}
    \label{fig:realisticSharpVS}
\end{figure}

We also investigated whether an atomically sharp cap or spike interface has a greater effect on valley splitting. Fig.~\ref{fig:realisticSharpVS} shows the valley splitting distributions as one of the interface widths $\sigma_{\mathrm{cap}}$ or \ $\sigma_{\mathrm{spike}}$ is varied while the other is kept atomically sharp. Across the full range of widths swept, keeping either interface sharp yields $E_v \gtrsim 1$~meV. However, a sharp cap with a spread-out 4\% spike gives a larger valley splitting than a sharp spike with a spread-out cap interface, indicating that the sharpness of the cap interface is more important.
\end{document}